\documentclass{ptephy_v1}%%%%%% to generate preprint number with ptep logo

\preprintnumber{XXXX-XXXX} %%% %%% Insert preprint number here

\def\rmF{{\rm F}}
\def\rmd{{\rm d}}
\def\om{\omega}

\def\half{{\textstyle \frac12}}

\newcommand{\lsim}{\stackrel{\scriptstyle <}{\phantom{}_{\sim}}}
\newcommand{\gsim}{\stackrel{\scriptstyle >}{\phantom{}_{\sim}}}

\begin{document}

\title{Condensate of excitations in moving superfluids}

%%%% To generate auto affiliation numbers please use \author{}\affil{} command

\author[1]{E.E. Kolomeitsev}
\affil{Matej Bel University, SK-97401 Banska Bystrica, Slovakia\email{Evgeni.Kolomeitsev@umb.sk}}

\author{D.N. Voskresensky}
\affil{National Research Nuclear University ``MEPhI'', 115409 Moscow, Russia \email{D.Voskresensky@gsi.de}
}

%%% To include the collaborator name... Please use the command "\collaborator"
%%% For example: \collaborator{ATLAS Collaboration}

\begin{abstract}%
A possibility of the condensation of excitations with a non-zero momentum in rectilinearly moving and rotating superfluid bosonic and fermionic (with Cooper pairing) media is
considered in terms of a phenomenological order-parameter functional at zero and non-zero temperature. The results might be applicable to the description of bosonic systems like superfluid $^4$He, ultracold atomic Bose gases, charged pion and kaon condensates in rotating neutron stars, and various superconducting fermionic systems with pairing, like proton and color-superconducting components in compact stars, metallic superconductors, and neutral fermionic systems with pairing, like the neutron component in compact stars and ultracold atomic Fermi gases. Order parameters of the ``mother'' condensate in the superfluid and the new condensate of excitations, corresponding energy gains, critical temperatures and critical velocities are found.
\end{abstract}

\subjectindex{D41,E32,I20,I61}

\maketitle

\section{Introduction}

A possibility of the condensation of rotons in the superfluid helium (He-II) moving in a capillary at zero temperature with a flow velocity exceeding the Landau critical velocity $v_{c}^{\rm L}$ was suggested in~\cite{Pitaev84}.  In~\cite{Voskresensky:1993uw} the condensation of excitations with a non-zero momentum in various relativistic and non-relativistic cold media moving with velocity exceeding $v_{c}^{\rm L}$  was studied further with the help of the effective Lagrangian for the complex scalar field, which describes Bose excitations in the medium. The Landau critical velocity is determined by the minimum of $\epsilon (k)/k$ at finite momentum $k$, where $\epsilon (k)$ is a branch of the spectrum of Bose excitations. Possible manifestations of the phenomenon in the bulk of He-II, rotating neutron stars with and without pion condensate, nuclei at high angular momentum and heavy-ion collisions were discussed.
Similar effect can occur also in a normal Fermi liquid with a zero-sound branch in the spectrum of particle-hole excitations \cite{Vexp95,Kolomeitsev:2015pia}. When the velocity of the Fermi liquid exceeds the Landau critical velocity related to this branch,
the number of excitations should grow exponentially with time and in the course of their interactions they may form a Bose condensate with a finite momentum. This possibility was studied in~\cite{Vexp95} for a moving Fermi liquid at finite temperature. Various consequences of the phenomenon in application to nuclear systems were announced.  In~\cite{Ancilotto05} the results of~\cite{Pitaev84} for He-II in a capillary were extended to He-II in a bulk. The condensation  of excitations in cold atomic Bose gases moving with a flow velocity exceeding $v_{c}^{\rm L}$  was  considered in~\cite{BP12}. A role of a Bose condensate of zero-sound-like excitations with non-zero momentum in the description of the stability of $r$ modes in rapidly rotating pulsars was discussed in~\cite{KV2015}.

Below, we study a possibility of the condensation of excitations in a state with a
non-zero momentum in moving media in the presence of a superfluid subsystem. The systems of our interest are neutral bosonic superfluids, such as the superfluid $^4$He, cf.~\cite{Pitaev84,Melnikovsky,LP1981,GS,Khalatnikov}, cold Bose atomic gases, cf.~\cite{BP12,coldatmode}, and inhomogeneous $\bar{K}^0$ condensates in neutron stars, cf.~\cite{Kolomeitsev:1995xz,Kolomeitsev:2002pg}, charged bosonic superfluids like $\pi^+$ and $\pi^-$ and $K^-$  condensates with $k\neq 0$ in neutron stars, cf.~\cite{Migdal:1990vm,Voskresensky:1993uw,Kolomeitsev:1995xz,Kolomeitsev:2002pg}; and various Fermi systems with the Cooper pairing, like the neutron superfluid in neutron star interiors, cf.~\cite{ST83}, cold Fermi atomic gases, cf.~\cite{Kagan}, neutron gas in neutron star crusts, cf.~\cite{Urban}, or charged superfluids, as paired protons in neutron star interiors, cf.~\cite{ST83}, paired quarks in color-superconducting regions of hybrid stars, cf.~\cite{Bailin:1983bm}, and paired electrons in metallic superconductors, cf.~\cite{Tilly-Tilly,Abrikosov}.

The key idea of the phenomenon  is the following ~\cite{Pitaev84,Voskresensky:1993uw}: When a medium moves as a whole with respect to a laboratory frame with a velocity higher than $v_{c}^{\rm L}$, it may become energetically favorable to transfer a part of its momentum from particles of the moving medium to a Bose condensate of  excitations (CoE) with a non-zero momentum $k\neq 0$. It would happen, if the spectrum of excitations is soft in some region of momenta. References~\cite{Pitaev84,BP12} studied the condensation of excitations at $T=0$ assuming the conservation of a flow velocity. Alternatively, we consider systems at other conditions, assuming the conservation of the momentum (or angular momentum for rotating systems) as in~\cite{Voskresensky:1993uw}. We consider bosonic and fermionic superfluid systems moving initially with the flow velocity above $v_{c}^{\rm L}$ both for $T=0$ and $T\neq 0$  (the latter case was not yet considered in mentioned references), taking into account a back reaction of the CoE on the ``mother'' condensate of the superfluid.

The work is organized as follows.
In Sect.~\ref{moving-fluid} we construct the phenomenological order-parameter functional for the description of  the CoE  coupled with the mother condensate in the superfluid moving linearly with the flow velocity exceeding $v_{c}^{\rm L}$. Section~\ref{Cold-superfluid}
is devoted to the description of cold moving superfluids.
Section~\ref{Warm-superfluid} studies peculiarities of the two-fluid motion in warm superfluids in the presence of the CoE. In Sect.~\ref{Vortices} we discuss a particular role of vortices. Some  numerical estimations valid for fermion superfluids in the BCS limit and for  He-II are performed in Sect.~\ref{Estimates}. Section~\ref{Pulsars} describes the CoE in rotating systems with application to the rapidly rotating pulsars. Section~\ref{Conclusion} contains concluding remarks.

\section{Order-parameter functional for moving fluid}\label{moving-fluid}

In the spirit of the Landau phenomenological theory of a second-order phase
transition the free-energy density of the superfluid subsystem in
its rest frame can be expanded in the order parameter $\psi$ for
temperatures $T\leq T_c$, where $T_c$ is the critical temperature
of the second-order phase transition, \cite{LP1981,GS}:
\begin{eqnarray}
F_{\rm L}[\psi]= c_T |\hbar\nabla\psi|^2/2 - a_T\,|\psi|^2 + b_T\,|\psi|^4/2\,.
\label{FGL}
\end{eqnarray}
Here $a_T\geq 0$, $b_T>0$ and $c_T>0$ are phenomenological
parameters depending on the temperature, so that $a_{T}$ vanishes at $T=T_c$.
When applied to superconductors the functional (\ref{FGL}) is
known in the literature as the Ginzburg-Landau model~\cite{LP1981}, while for the case of the superfluid $^4$He it is called the Ginzburg-Pitaevskii model. The phenomenological
description of cold weakly interacting Bose gases was performed by
Gross and Pitaevskii, see~\cite{LP1981}. As pointed out in ref.~\cite{GS}, the expansion in the order parameter is a primary
feature in the Landau's phase-transition theory, whereas an
expansion in powers of $(T_c - T)$ is a secondary assumption.
Therefore, we will use the functional (\ref{FGL}) for all $T<T_c$.

For $0<t=1-T/T_c\ll 1$, the coefficients $a_T$ and $b_T$ can be
expanded as~\cite{GS} $a_T=a_0\,t^\alpha$\,, $b_T=b_0\,t^\beta
$\,, and $c_T$ is usually assumed to be constant, $c_T=c_0$.
Within the mean-field approximation from the
Taylor expansion of $F_{\rm L}$ in $t\ll 1$ it follows that
$\alpha =1$, $\beta =0$. The width of the  fluctuation
region, wherein the mean-field approximation is not applicable,
is evaluated with the help of the Ginzburg~\cite{GS} and Ginzburg-Levanyuk~\cite{GL} criteria.
For the ordinary metallic superconductors the fluctuation region
proves to be usually very narrow and the mean-field approximation
holds then for almost any temperatures below $T_c$, except a tiny
vicinity of $T_c$. Thus, for $t\ll 1$, neglecting the mentioned narrow
fluctuation region, one may use $\alpha =1$, $\beta =0$.
For He-II, fluctuations prove to be important for all temperatures
below $T_c$, cf.~\cite{GS}.
Using the experimental fact that the specific heat of the
He-II has no power divergence at $T\to T_c$, we get
$\alpha =4/3$ and $\beta =2/3$ that coincides with phenomenological
findings~\cite{GS}.

%%%%%%%%%%%%%%%%%%%%%%%%%%%%%%%%%%%%%%%%%%%%%%%%%%
%%%%%%%%%%%%%%%%%%%%%%%%%%%%%%%%%%%%%%%%%%%%%%%%%%

Consider a system at a finite temperature consisting of normal and superfluid parts undergoing  rectilinear motions parallel to a wall.
The wall singles out the laboratory frame with respect to which the motion is defined.
Interactions between particles in normal fluid may lead to creation of excitations.
Mechanisms of the excitation production depend on the specifics of problems and will be discussed below in Sects.~\ref{Warm-superfluid}, \ref{Vortices}, and \ref{Pulsars}.

We assume that the superfluid  moves  with an initial velocity $\vec{v}$ with respect to the wall and additionally the excitations can carry some  net momentum, $\vec{j}_n$, with respect to the superfluid. Then one can define an average velocity of the excitations with respect to the superfluid component $\vec{w}$. With respect to the wall the excitations have the average velocity $\vec{v}_n=\vec{w}+\vec{v}$. The motion of the superfluid as a whole with velocity $\vec{v}$  relative to the reference frame of the wall can be described by introducing  the phase of the condensate field
$\psi=|\psi|e^{i\phi}$ with $\vec{v}=\hbar\nabla\phi/m\,.$

We can write the variational functional for the condensate field in the standard form of the two-fluid model \cite{Khalatnikov}
\begin{eqnarray}
F[\psi,\vec{v},\vec{v}_n]=\frac12\rho_s \vec{v}\,^2 + \frac12\rho_n\vec{v}_n^{\,2} + {F}_{\rm bind} + F_{\rm L}[\psi].
\label{F2fluids}
\end{eqnarray}
The density of the superfluid component, which determines the amplitude of the condensate field $\psi$ is related to the normal component $\rho_n$ by the relation
\begin{eqnarray}
m\,|\psi|^2=\rho_s(T,\vec{w})=\rho-\rho_n(T,\vec{w})\,,
\quad \rho_n(T,\vec{w})=(\vec{j}_n\vec{w})/\vec{w}\,^2\,,
\label{rhos-def}
\end{eqnarray}
where $m$ is the mass of the pair for systems with pairing, and the mass of a boson in
bosonic superfluids, e.g., the  mass of the $^4$He atom in
case of the He-II. The quantity ${F}_{\rm bind}$ in Eq.~(\ref{F2fluids}) stands for a binding free-energy density of the normal subsystem in its rest
frame, which explicit form is not of our interest here.
The first term in (\ref{F2fluids}) can be hidden in $F_L[\psi]$ as a phase of the condensate field. For the case when the normal component rests, $v_n=0$, i.e., the superfluid moves with the velocity $\vec{v}=-\vec{w}$, the minimization of the functional (\ref{F2fluids}) gives
\begin{eqnarray}
|\psi_{\rm eq}(w)|^2=(a_T-m\,w^2/2)/b_T\,,
 \label{psiw}
\end{eqnarray}
and, hence, the critical temperature decreases with a velocity increase as  $T_c(w)=T_c(1-mw^2/(2a_0))$~\cite{Amit68}
and vanishes at $w=w_{\rm A}=(2a_0/m)^{1/2}$.
In reality the superfluid flow $\vec{j}_s=\vec{v}\rho_s=\vec{v}\rho-\vec{j}_n$ becomes unstable with $w\neq 0$
even at the smaller velocity $w_{\rm A1}$, determined from the
condition $\partial j_s(T,w)/\partial {w} =0$, see~\cite{Kramer,Andreev-Melnik04}. In general $w_{\rm A}$ is smaller than $v_c^{\rm L}$~\cite{Kramer} and for small $t$ one finds~\cite{GS,Kramer} $w_{\rm A1}\approx (2\,a_T/$ $({3}m))^{1/2}\ll v_c^{\rm L}$.
Thus, for a flow in a narrow pipe, in the equilibrium state with $v_n=0$ and hence $w=v$, the CoE would not appear since the mother condensate is destroyed already for $v=w_{\rm A1}<v_c^{\rm L}$.
Therefore, in further discussion we assume that $w<w_{\rm A1}$. Situations, in which the latter condition is fulfilled, will be discussed later in the text. In case $w<w_{\rm A1}$ of our interest the finite value of $w$ implies only a redefinition of the critical temperature $T_c\to T_c(w)$. Thereby, to simplify further notations we  put $w=0$. The generalization is straightforward.
Then the free-energy density functional of the system moving with the velocity $v$ respectively the wall is given by
\begin{eqnarray}
F[\psi,\vec{v}\,] = {\rho v^2}/{2}+\bar{F}_{\rm bind} + F_{\rm L}[\psi]\,.
\label{Fw=0}
\end{eqnarray}
The equilibrium volume-averaged value of the condensate is given then by Eq.~(\ref{psiw}) and the volume-averaged density of the normal component, $\bar{\rho}_n$, is related to the averaged total density of the fluid, $\bar\rho$, as  $\bar\rho_n =\bar\rho - m\,|\psi^{\rm eq}|^2$.
The equilibrium value of the volume-averaged free-energy density
(we shall call it as an ``in''-state) is
\begin{eqnarray}
\bar{F}_{\rm in}=\bar{\rho}v^2/2 + \bar{F}_{\rm bind} -a_T^2/(2b_T)\,.
\label{Fin}
\end{eqnarray}

When the speed of the flow $v$ exceeds the Landau critical
velocity,
$$v_c^{\rm L}=\min_{k}(\epsilon(k)/k)\equiv\epsilon(k_0)/k_0\,,$$
on top of the mother condensate $\psi$ there may appear in the fluid a CoE $\psi'$~\cite{Pitaev84,Voskresensky:1993uw,BP12} with
the frequency $\epsilon(k_0)$ and momentum $k_0$ calculated in the rest frame of the
superfluid, where, as we have assumed, the  ratio $\epsilon (k)/k$ has
minimum at $k=k_0\neq 0$. For He-II the spectrum $\epsilon(k)$ is the standard
phonon-roton spectrum, normalized as $\epsilon(k)\propto k$ for
small $k$.
In the case of the straightforward motion, we, following the symmetry arguments, may choose the simplest form of the CoE order parameter depending on the time $\tau$ and the coordinate $\vec{r}$ as
\begin{eqnarray}
\label{ex}
\psi'=\psi'_0 e^{-i(\epsilon (k_0)\tau - \vec{k}_0\vec{r})/\hbar}
\end{eqnarray}
with a constant amplitude $\psi'_0$ for the homogeneous system
that we consider.

For the description of CoE with the given
frequency $\epsilon(k_0)$ the functional (\ref{FGL}) must be
supplemented by the functional $F_{\rm ex}[\psi]$ involving higher
gradient terms so that the variation of the Fourier transform
of the full functional reproduces the excitation frequency
$$
\epsilon(k_0)=\frac{\delta^2(F_{\rm L}[\psi+\psi']+F_{\rm ex}[\psi+\psi'])}{\delta\psi'\delta\psi'^*}\Bigg|_{\psi'=0}
$$
and the  self-interaction parameters of the CoE free-energy density functional:
\begin{eqnarray}
2b'_{T,k_0}&=&
\frac{\delta^4(F_{\rm L}[\psi+\psi']+F_{\rm ex}[\psi+\psi'])}{\delta{\psi}\delta{\psi}^*\delta{\psi'}\delta{\psi'}^{*}}\Bigg|_{
\psi'=0} ,
\nonumber\\
2b''_{T,k_0}&=&\frac{\delta^4(F_{\rm L}[\psi+\psi']+F_{\rm ex}[\psi+\psi'])}{\delta{\psi'}^2\delta{\psi'}^{*2}}\Bigg|_{
\psi'=0}\,.
\nonumber
\end{eqnarray}
For example, in  ref.~\cite{BP12} these parameters were estimated for a cold weakly interacting Bose gases. The explicit structure of $F_{\rm ex}$ is not important for our study as we use the phenomenological parameters $b'_{T,k_0}$, and $b''_{T,k_0}$.

We suppose that, when the CoE is formed (we shall call it a ``fin''-state), the
initial momentum density is redistributed between the fluid and
the CoE:
\begin{eqnarray}
\label{momentumcons}
\bar{\rho}\, \vec{v}=(\bar{\rho}- m\,|\psi'_0|^2)\, \vec{v}_{\rm fin} + (\vec{k}_0+m\vec{v}_{\rm
fin})\, |\psi'_0|^2\,.
\end{eqnarray}
Here $\vec{k}_0|\psi'_0|^2$ is the momentum density carried by the CoE in the rest frame of the superfluid, $(\vec{k}_0+m\vec{v}_{\rm fin})|\psi'_0|^2$
is the resulting momentum density carried by the CoE in the laboratory frame and the first term, $(\bar{\rho}- m\,|\psi'_0|^2)\, \vec{v}_{\rm fin}$, is the resulting momentum density carried by the superfluid in the laboratory frame. So, the CoE necessarily moves in the laboratory frame.

In the presence of the CoE the resulting order parameter
$\psi_{\rm fin}$ is the sum of the mother condensate, $\psi$, and of the CoE, $\psi'$,
$\psi_{\rm fin} =\psi +\psi'$\,. The volume-averaged free-energy density
of the system with the CoE, $\bar{F}_{\rm
fin}=\bar{F}_{\rm L}[\psi_{\rm fin}]+\bar{F}_{\rm ex}[\psi_{\rm
fin}]$, can be written as
\begin{eqnarray}
&&\bar{F}_{\rm fin}[\psi,\psi'] = {\textstyle \frac12}\bar{\rho}\,v_{\rm fin}^2
+ \bar{F}_{\rm bind} -a_T\,|\psi|^2+ \half b_T\,|\psi|^4
\\
&&\quad+ (\tilde{\epsilon}(k_0)-a_T)\,|\psi'|^2 + 2\,b'_{T,k_0}
|\psi|^2 |\psi'|^2 + \half b''_{T,k_0} |\psi'|^4\,,
\nonumber
\end{eqnarray}
where $\tilde{\epsilon}(k)$ is the energy of the excitation including the mean-field
potential, $\tilde{\epsilon}(k)=\epsilon(k) + a_T(1-2 b'_{T,k_0}/b_T)$.
%%%%%%%%%%%%%%%%%%%%%%%%%%%%%%%%%%%%
Now, using the momentum conservation (\ref{momentumcons}) we
express $\vec{v}_{\rm fin}$ through $\vec{v}$ and get for the
change of the averaged free-energy density associated with the
 CoE,
\begin{eqnarray}
&&\delta \bar{F}[\psi,\psi'] =\half b_T\big(|\psi|^2- a_T/b_T\big)^2 +k_0\,\big(v_c^{\rm L} - v\big)|\psi'|^2
\nonumber\\ &&\,\,+ 2 b'_{T,k_0}\, \big(|\psi|^2-a_T/b_T\big)\,|\psi'|^2 + \half
(b''_{T,k_0}+k_0^2/\bar{\rho})|\psi'|^4\,,
\nonumber\\
\label{Fgen}
\end{eqnarray}
where we put
$\vec{k}_0\parallel \vec{v}$\,. We apply now the functional (\ref{Fgen}) to superfluids for $T\to 0$  and  $T\neq 0$.

\section{Cold  superfluid}\label{Cold-superfluid}

\subsection{bosonic system}

At $T\to 0$ the whole medium is superfluid and amplitudes of the condensates are constrained by the spatially averaged particle number density
\begin{eqnarray}
\label{n}
\bar{n}=\overline{|\psi+\psi'|^2}=|\psi|^2+|\psi'|^2 \,.
\end{eqnarray}
In the presence of the CoE the density becomes spatially oscillating around its averaged value.  For a weak condensate, {\it i.e.}, $|v-v_c^{\rm L}|\ll v_c^{\rm L}$, we find perturbatively
\begin{eqnarray}
\delta n=n-\bar{n}\approx 2\sqrt{n}|\psi'_0|\cos((\epsilon(k_0)\,\tau
-\vec{k}_0\vec{r})/\hbar\,).
\end{eqnarray}
The density modulation was predicted in~\cite{Pitaev84} and reproduced in the numerical simulation of the supercritical flow in He-II using a realistic density functional~\cite{Ancilotto05}.

Replacing Eq.~(\ref{n}) in ~(\ref{Fgen}) and putting $T=0$ we
find the change of the spatially-averaged energy density of the system because
of the appearance of the CoE, $\delta
\bar{E}=\bar{E}_{\rm fin}-\bar{E}_{\rm in}$,
\begin{eqnarray}
\delta \bar{E}= k_0(v_c^{\rm L} - v)|\psi'_0|^2 +
k_0^2(1-\chi_0)|\psi'_0|^4/(2\bar{\rho})\,,
\end{eqnarray}
where $\chi_0=(4b'_{0,k_0}-b_0 -b''_{0,k_0})\bar{\rho}/k_0^2$\,, $b'_{0,k_0}, b_0, b''_{0,k_0}$ are considered above coefficients taken now for $T=0$.
Minimizing this functional with respect to $\psi'_0$ we obtain
\begin{eqnarray}
|\psi'_0|^2=\frac{\bar{\rho}\,(v-v_c^{\rm L})} {k_0\,(1-\chi_0)}\,\theta(v-v_c^{\rm L})\theta(1-\chi_0) \,. \label{T0-sol}
\end{eqnarray}
From~(\ref{momentumcons}) we find that because of the CoE  with $k\neq 0$ the flow is decelerated to the
velocity
\begin{eqnarray}
v_{\rm fin}=v_c^{\rm L}-(v-v_c^{\rm
L})\chi_0/(1-\chi_0)\,.\label{vfin0}
\end{eqnarray}
 The volume-averaged energy gain due to
appearance of the CoE is
\begin{eqnarray}\label{deltaE}
\delta \bar{E}= -\frac{\bar{\rho}\,(v-v_c^{\rm L})^2}
{2\,(1-\chi_0)}\theta(v-v_c^{\rm L})\,.
\end{eqnarray}
If ${\chi}_0>0$, one has $v_{\rm fin}<v_c^{\rm L}$. As we estimate below in case of He-II and in case of the BCS weak coupling, the parameter $|\chi_0|\ll 1$ and $v_{\rm fin}\simeq v_c^{\rm L}$.

As follows from Eq.~(\ref{deltaE}) the CoE appears in a second-order phase transition since
$\frac{\rmd \delta\bar{E}}{\rmd v}\Big|_{v_c^{\rm L}}=0$ but $\frac{\rmd^2 \delta\bar{E}}{\rmd v^2}\Big|_{v_c^{\rm L}}\neq 0$. The amplitude of the CoE~(\ref{T0-sol}) grows with the velocity, whereas the amplitude of the mother condensate decreases. The value $|\psi|^2 $ vanishes when $v=v_{c2}$, the second critical velocity, at which $|\psi'_0|^2=\bar{n}$ according to Eq.~(\ref{n}). The value $v_{c2}$ is evaluated from~(\ref{T0-sol}) as
$$v_{c2}=v_c^{\rm L}+k_0(1-\chi_0)/m\,.$$
When the mother condensate disappears at $v=v_{c2}$, the excitation spectrum is cardinally reconstructed, and the superfluidity destruction occurs as a first-order phase transition. We assume that for $v>v_{c2}$ the excitation spectrum has no low-lying local minimum at a finite momentum. Then the amplitude $|\psi'_0|^2$ jumps from  $\bar{n}$ to 0 and $\delta \bar{E}$ jumps from $\delta \bar{E}(v_{c2})=-\bar{\rho} k_0^2(1-\chi_0)/(2m^2)$ to 0 at $v=v_{c2}$.

%%=======================================

\subsection{fermionic system}

As shown in refs.~\cite{Kagan,Urban,Kulik43}, in fermionic systems with pairing there may exist bosonic modes with suitable spectra,  supporting quasiparticle excitations
with the energy $\simeq 2\Delta$ and momentum  $k_0 \simeq 2p_{\rm F}$, $\Delta$ is the pairing gap computed in the rest frame of the superfluid, see Fig.~2 in~\cite{Kagan},  and Fig.~4 in~\cite{Urban}. For these modes the Landau critical velocity is \begin{eqnarray}
v_c^{\rm L} \simeq \Delta/p_{\rm F}\,,
\label{vcL-bos}
 \end{eqnarray}
and for $v> v_c^{\rm L}$ there is a chance for the condensation of the bosonic  excitations as we considered above.

Besides bosonic excitations there exist
fermionic ones with the spectrum $\epsilon_{\rm
f}(p)=\sqrt{\Delta^2+v_{\rm F}^2(p-p_{\rm F})^2}$\,. Stemming from
the breakup of Cooper pairs, the fermionic excitations are
produced pairwise and the corresponding (fermion) Landau
critical velocity is
$v_{c,{\rm f}}^{\rm
L}=\min_{\vec{p}_1,\vec{p}_2} [(\epsilon_{\rm
f}(p_1)$ $+\epsilon_{\rm f}(p_2))/|\vec{p}_1+\vec{p}_2|]$\,. The
latter expression reduces to~\cite{Castin14}
\begin{eqnarray}
v_{c,{\rm f}}^{\rm L} = (\Delta/p_{\rm F})/(1+\Delta^2/p_{\rm F}^2 v_{\rm
F}^2)^{1/2}\,.
\end{eqnarray}
We see that up to a small correction of the order of $(v^{\rm L}_c /v_{\rm
F})^2$ $\ll 1$, $v_{c,{\rm
f}}^{\rm L}\simeq v_c^{\rm L}$. More accurately we get  $v_c^{\rm L}-v_{c,{\rm f}}^{\rm L}\approx\frac12 v_c^{\rm L} (v_c^{\rm L}/v_\rmF)^2$.

For $T\to 0$ the fermionic excitations are produced near the wall  and move, therefore, with respect to the superfluid with the velocity $-\vec{v}$.\footnote{At finite temperatures fermionic excitations are mainly produced inside the pre-existing normal component moving with the velocity $\vec{w}$ with respect to the superfluid component.} Hence, the change of the energy density due to the Cooper pair breaking can be calculated as
\begin{eqnarray}\label{deltaEf}
\delta \bar{E}_{\rm pair} &=& \intop\frac{2\rmd^3p}{(2\pi\hbar)^3}({\epsilon_{\rm
f}(p)-\vec{p}\,\vec{v}}\,)\theta (\epsilon_{\rm f}(p)-\vec{p}\,\vec{v}\,)\,.
\end{eqnarray}
Expanding this integral for velocities $v$ close to the critical velocity $v_c^{\rm L}\approx v_{c,{\rm f}}^{\rm L}$ we find
\begin{eqnarray}\label{deltaEf2}
\delta \bar{E}_{\rm pair}
 &\approx& - 2\sqrt{2}\bar{\rho} (v_{c}^{\rm L})^{-1/2}(v- v_{c,{\rm f}}^{\rm L})^{5/2}\,,
 % [v/v_{c}^{\rm L}-1]^{1/2}\,.
\end{eqnarray}
being valid for $v\ll v_{\rm F}$. Since the critical velocity $v_{c,{\rm f}}^{\rm L}$ is slightly smaller than $v_c^{\rm L}$, Eq.~(\ref{deltaEf2}) wins over Eq.~(\ref{deltaE}) for $v=v_c^{\rm L}$, but already for the velocities
$v>v_c^{\rm L}[1+(v_c^{\rm L}/v_\rmF)^{5/2}]$ the formation of the CoE becomes energetically more favorable than the pair breaking. Although the above estimates are applicable only for $0<v/v_{c}^{\rm L}-1\ll 1$, there is another argument in favour of the condensation of bosonic excitations. In a system, in which the normal component (fermionic excitations) moves relative to superfluid with the velocity $w$ the pairing gap decreases (Rogers-Bardeen effect~\cite{Bardeen}).
In the case under consideration a superfluid moves with the velocity $v>v_c^{\rm L}$ relative to the wall. Excitations are produced near the wall, and the pairing gap  decreases,  being determined by the equation~\cite{Zagozkin}
\begin{eqnarray}
\ln\frac{p_\rmF v}{\Delta}=\Big(1-\frac{\Delta^2(v)}{p_\rmF^2 v^2}\Big)^{1/2}
-\ln\Big(1+\sqrt{1-\frac{\Delta^2(v)}{p_\rmF^2 v^2}}\Big).\label{Gap-vv}
\end{eqnarray}
For $0\le v/v_c^{\rm L}-1\ll 1$ this equation has the solution
\begin{eqnarray}
\Delta(v)/\Delta\approx 1-(3/2)(v/v_c^{\rm L}-1)^2\,.
\label{Gap-v}
\end{eqnarray}
With the subsequent growth of $v$ (for $v/v_c^{\rm L}-1\gsim 1$) the gap continues to decrease and, as follows from Eq.~(\ref{Gap-vv}), it vanishes
at $v=v_{c2,{\rm f}}^{\rm L}=\frac{e}{2}v_c^{\rm L}$,  see~\cite{Zagozkin}.
Since in the presence of the CoE the final velocity of the flow is $v_{\rm fin}= v_{c}^{\rm L}$ and the gap does not change, the additional gain in the energy density due to the formation of the condensate of bosonic excitations compared to the pair breaking without the CoE formation is
\begin{eqnarray}
\delta \bar{E}_{\rm gap} &=& F_{\rm L}^{\rm eq}(T=0,\Delta)-F_{\rm L}^{\rm eq}(T=0,\Delta(v))\,,
\quad
\label{deltaEgap}
\end{eqnarray}
where~\cite{LP1981}
$F_{\rm L}^{\rm eq}(T=0,\Delta)=-\frac{m^*p_\rmF}{4\pi^2}\Delta^2$.
For $0\le v/v_c^{\rm L}-1\ll 1$ by substituting Eq.~(\ref{Gap-v}) in Eq.~(\ref{deltaEgap}) and rewriting $\frac{m^*p_\rmF}{4\pi^2}\Delta^2=\frac34\rho (v_c^{\rm L})^2$ we easily find
\begin{eqnarray}
\delta \bar{E}_{\rm gap}\approx -(9/8)\bar{\rho} (v-v_c^{\rm L})^2\,.
\end{eqnarray}
For $v>v_{c2,{\rm f}}^{\rm L}$ one has $\Delta (v)=0$, and, as follows from Eq.~(\ref{deltaEgap}), the gain in the energy density because of the CoE compared to the full destruction of the pairing would be $\delta\bar{E}_{\rm gap}=-3\bar{\rho} ( v_c^{\rm L})^2/4$.

Thus we can conclude that the creation of the condensate of bosonic excitations with finite momentum in moving cold fermionic systems with pairing leading to a reduction of the flow velocity is energetically more profitable than the breaking of Cooper pairs and the decrease of the pairing gap.

\section{Warm superfluid. Two-fluid motion}\label{Warm-superfluid}

Only for a very low $T$ the normal component can be neglected.
For a higher temperature the normal subsystem serves as a
reservoir of particles for the formation of the mother and daughter
condensates, which amplitudes are now to be chosen by minimization
of the free energy of the system. Therefore,
minimizing~(\ref{Fgen}), we vary now $\psi$ and $\psi'_0$
independently and find
\begin{eqnarray}
&&|\psi'_0|^2 =\frac{\bar{\rho} \big(v-v_c^{\rm
L}\big)}{k_0(1-{\chi}_T)} \theta\big(v-v_c^{\rm
L}\big)\,\theta\big(1-{\chi}_T\big) \,,
\label{orderp}\\
&&|\psi|^2 = \Big(\frac{a_T}{b_T}  -
2\frac{b'_{T,k_0}}{b_{T}}|\psi'_0|^2\Big)\, \theta
(\widetilde{T}_c(v) -T)\theta (v_{c2}(T)-v)\,, \nonumber
\end{eqnarray}
where ${\chi}_T=(4b'^2_{T,k_0}/b_T-b''_{T,k_0})\bar{\rho}/k_0^2$. The
quantity $\widetilde{T}_c$ stands for the renormalized critical
temperature, which depends now on the flow velocity, and
$v_{c2}(T)$ stands for the second critical velocity depending on
$T$. The condition $|\psi|^2=0$ implies the relation between $v$ and $T$,
\begin{eqnarray}\label{tildet}
v=v_c^{\rm L} + a_T k_0(1-{\chi}_T)/{(2b'_{T,k_0}\bar{\rho})}\,.
\end{eqnarray}
The solution of this equation for the velocity, $v_{c2}(T)$,
increases with the decreasing temperature, and the solution for
the temperature, $\widetilde{T}_c (v)$, decreases with increasing
$v$. At $T=\widetilde{T}_c (v)$ or $v=v_{c2}(T)$ we have
$|\psi|^2=0$ but $|\psi_0^{'}|^2\neq 0$, and for
$T>\widetilde{T}_c (v)$ or for $v>v_{c2}(T)$ the condensate
$|\psi_0^{'}|^2$ vanishes, as for $|\psi|^2=0$ the
spectrum of excitations does not contain a suitable low-lying branch. Thus,
the superfluidity is destroyed at $T=\widetilde{T}_c (v)$ or
$v=v_{c2}(T)$ in a first-order phase transition.

From Eqs.~(\ref{momentumcons}) and (\ref{orderp}) we find for
$v>v_c^{\rm L}$ and ${\chi}_T < 1$ the resulting velocity of the
flow
\begin{eqnarray}
\label{vfin}
v_{\rm fin} = v_c^{\rm L} -  (v - v_c^{\rm L}){\chi}_T/(1-{\chi}_T)\,,
\end{eqnarray}
similar to Eq.~(\ref{vfin0}) obtained above for $T=0$.
If ${\chi}_T>0$, one has $v_{\rm fin}<v_c^{\rm L}$, and $v_{\rm
fin}\simeq v_c^{\rm L}$ for $0<\chi_T \ll 1$.

Substituting the order parameters from (\ref{orderp}) in
(\ref{Fgen}), we find for the averaged free-energy density gain
owing to the appearance of the CoE
\begin{eqnarray}
\label{deltaF}
\delta\bar{F} =-\half\bar{\rho}(v-v_c^{\rm L})^2(1-{\chi}_T)^{-1}
\theta(v - v_c^{\rm L})\, \theta(v_{c2}-v)
\end{eqnarray}
for ${\chi}_T<1$\,. Thus, for $v_c^{\rm L}<v<v_{c2}$ the free
energy decreases owing to the appearance of the CoE with $k\neq 0$
in the presence of the non-vanishing mother condensate. The value
of $k_0$ is to be found from the minimization of
Eq.~(\ref{deltaF}). As $\widetilde{T}_c$, the momentum $k_0$ gets
renormalized and differs now from the value corresponding to the
minimum of $\epsilon (k)/k$. As for $T=0$, for $T\neq 0$ the CoE
appears at $v=v_c^{\rm L}$ in a second-order phase transition but
it disappears at $v=v_{c2}$ in a first-order phase transition with
jumps from
\begin{eqnarray}
\label{jumpvc2}
{\delta \bar{F}(v_{c2})} = -\frac{a^2_T
k_0^2}{8b'^2_{T,k_0}\bar{\rho}}(1-{\chi}_T),\,\,\,
|\psi'_0(v_{c2})|^2=\frac{a_T}{2b'_{T,k_0}}\,
\end{eqnarray}
to 0.

At finite temperature the dynamics of the CoE amplitude can be determined from the equation \cite{Lif81}
\begin{eqnarray}
\dot{\psi'_0} = -\Gamma\frac{\delta ( \delta \bar{F})}{\delta \psi'^{*}_0}\,,
\end{eqnarray}
where $\Gamma$ is a formation rate of the CoE. In the theory of non-equilibrium superconductors this equation is  known as the time-dependent Ginzburg-Landau equation. Note that the dynamics following this equation is different from that follows from the Gross-Pitaevskii equation describing a weakly non-ideal Bose gas in an external field. It is determined by the time-dependence of the potential.
We emphasize that the above consideration assumes that the formation rate $\Gamma$ of the CoE is faster than the deceleration rate
$1/\tau^{\rm norm}_{\rm fr}$ of the normal subsystem. The former time  $1/\Gamma$ is of a microscopic origin, whereas $\tau_{\rm fr}^{\rm norm}$ might be very large as being caused by the friction force between the normal component and the wall. For  rotating compact stars $\tau^{\rm norm}_{\rm fr}$ is determined by the decay of a star magnetic field yielding $\tau_{\rm fr}^{\rm norm}\gsim 10^3$\,yrs \cite{ST83} for magnetic fields below $10^{13}$\,G.
Thus, the COE has enough time to be developed in mentioned cases.

When the fluid flowing with $v>v_c^{\rm L}$ at
$T>\widetilde{T}_c (v)$ is cooled down to $T<\widetilde{T}_c (v)$,
it consists four components: the normal excitations, the
superfluid, the vortices and the CoE, all moving rigidly with $v_{\rm
fin}<v_c^{\rm L}$ (if ${\chi}_T>0$). If the system is then rapidly
re-heated to $T>\widetilde{T}_c (v)$, the superfluid component, the vortices and the
CoE vanish and the remaining normal fluid consists of two
fractions: one still moving with $v_{\rm fin}(\widetilde{T}_c)<v_c^{\rm
L}$, owing to conservation of the momentum,  and the other one, being originated from the melted CoE,
with the mass equal to $m a(\widetilde{T}_c)/(2b'_{T,k_0}(\widetilde{T}_c))$,
moving with a higher velocity until a new equilibrium is established. This may show one of possibilities how one could identify formation of the CoE experimentally.

Note that for fermion superfluids at $T\neq 0$  after the CoE is formed the flow velocity $v_{\rm fin}< v_{c,{\rm f}}^{\rm L}$, for $v-v_{c}^{\rm L}>4tv_{c}^{\rm L}/9$ (the estimate is done for
${\chi}_T =3b_0\bar{\rho}/k_0^2$), and hence the Cooper pair breaking does not occur, whereas the condensate of Bose excitations is preserved.

\section{Vortices}\label{Vortices}
%Below
Above we focused our consideration on the cases where either the
vortices are absent (as in a narrow capillary \cite{Pitaev84}) or they leave the system (in  open systems), or the presence of vortices supports a
common rigid motion of the normal and superfluid components \cite{Tilly-Tilly} (e.g., as in systems with charged components \cite{Sauls89}, or in rotating systems, like neutron stars \cite{ST83}).

In case of He-II moving in a narrow capillary
vortices do not appear, see~\cite{Pitaev84,Ancilotto05}. For a
rectilinearly moving superfluids in extended geometry there may
appear excitations of the type of vortex rings and other structures~\cite{AndreevKagan84}.
The energy of the ring is estimated~\cite{GS,Khalatnikov} as $\epsilon^{\rm vort}=2\pi^2
\hbar^2|\psi|^2 R$ $m^{-1} \ln (R/\xi)$, and the momentum is $p^{\rm vort}=2\pi^2\hbar|\psi|^2 R^2$, where $R$ is the radius of the vortex ring and $\xi$ is the coherence length, $\xi\sim\hbar(c_T/a_T)^{1/2}$, as estimated above. Thus, $v_{c1}=\epsilon^{\rm vort}/p^{\rm vort}= \hbar (R_t\,m)^{-1}\ln(R_t/\xi)$ is the Landau critical velocity for the vortex production, where
 in the absence of impurities $R_t$ is of the order of the transverse size of the system. For a system of distributed impurities moving together with the fluid, $R_t$ is a typical distance between the defects. Vortices are pined to the impurities and move together with them and the superfluid. In an open clean system at $v>v_{c1}$ the vortex rings are pushed
to infinity by Magnus and Iordanskii forces. Note that for spatially extended systems the value $v_{c1}$ is lower than the Landau critical velocity $v_c^{\rm L}$. The flow moving with the
velocity $v$ for $v_{c1}\leq v$ may be considered as metastable,
since the vortex creation probability is hindered by a large
potential barrier and formation of a vortex takes a long
time~\cite{vortex-creation}. The vortex production rate increases, however, strongly when $v$ approaches $v_c^{\rm L}$~\cite{vortex-creation}.
For a motion in a pipe the vortices are captured by the pipe wall,
forming after a while a stationary subsystem in the frame of the
walls. Periodic solitonic solutions of the Gross-Pitaevskii
equation were studied in~\cite{Tsuzuki}. This situation  might be
rather similar to that of a mother condensate moving in a periodic
potential, produced by the spatial variations of the CoE order
parameter~\cite{BP12}. Since in exterior regions of the
vortices the superfluidity persists, our
consideration of the condensation of excitations for $v_{c}^{\rm L}<v$ is applicable.
Note that in He-II under a high external pressure $v_{c}^{\rm L}$ decreases and at some conditions becomes lower than $v_{c1}$, see~\cite{McClintock}, and in the interval $v_{c}^{\rm L}<v<v_{c1}$ there are no vortices but the CoE may appear.

In superconducting systems vortices if formed, are involved in a common motion with the superconducting subsystem due to the appearance of a tiny London field~\cite{Sauls89} distributed throughout the medium, that supports the condition $w=0$.

In rotating superfluids vortices appear at rotation frequency
$\Omega >\Omega_{c1}= \frac{\hbar}{m\,R^2}\ln(R/\xi)$, where for the spherical system $R$ is the size of the system (transversal size for the cylindrical system), and their number grows with an increase of $\Omega$.
When the density of vortices becomes sufficiently large, they form a lattice, cf.~\cite{Tilly-Tilly}, forcing, thereby, the superfluid and normal components to move as a rigid body, i.e. with $w\to 0$.

\section{Estimates for fermionic and bosonic superfluids}\label{Estimates}

We apply now the expressions derived in the previous sections to several practical cases.

\subsection{fermionic syperfluid}

%First, we
Consider a fermion system with
%pairing.
%For the fermionic systems with
the singlet pairing.  In the weak-coupling
(BCS) approximation the parameters of the functional (\ref{FGL}) can be extracted from the
microscopic theory~\cite{LP1981}:
\begin{eqnarray}
c_0={1}/{2m^*_{\rm F}}\,, \,\, a_0 ={6\pi^2 T_c^2}/{(7\zeta(3)\mu)}\,, \,\, b_0={a_0}/{n}\,,
\label{BCS-param}
\end{eqnarray}
where $m^*_{\rm F}$ stands for the effective fermion mass
($m^*_{\rm F}\simeq m_{\rm F}$ in the weak-coupling limit),
$n=p_{\rm F}^3/(3\pi^2\hbar^3)$ is the particle number density,
and the fermion chemical potential is $\mu\simeq \epsilon_{\rm F}=
{p_{\rm F}^2}/{(2m^{*}_{\rm F})}$. The function $\zeta(x)$ is the
Riemann $\zeta$-function and $\zeta (3)=1.202$.
With the BCS parameters we have $|\psi |^2=nt$ and the pairing gap
$\Delta = T_c\sqrt{\frac{8\pi^2 t }{7\zeta (3)}}$,
see~\cite{Abrikosov}.

With  parameters~(\ref{BCS-param}) we estimate $b_0\bar{\rho}/k_0^2 = 3\Delta^2/(8v_{\rm F}^2p_{\rm F}^2)$ and
$a_0/k_0=3\Delta^2/(4v_{\rm F} p_{\rm F}^2)\,, $ where $\bar{\rho}
\simeq \bar{n}m_{\rm F}$. We see that if $b'_{T,k_0}\sim
b''_{T,k_0}\sim b_T$ one gets $0<{\chi}_T=3b_T\bar{\rho}/k_0^2\ll 1$, since the latter
inequality is reduced to the inequality $\Delta \ll \epsilon_{\rm
F}$, which is well satisfied. In this limit $|\psi_0'|^2$ given by Eq.~(\ref{orderp}) gets the same form as Eq.~(\ref{T0-sol}). The resulting flow velocity after condensation of
excitations,~(\ref{vfin}), is lower than $v_{c}^{\rm L}$  but close to it.

Since for the BCS case we have $\alpha =1$, $\beta =0$, Eq.~(\ref{tildet})
for the new critical temperature is easily solved, for $v>v_c^{\rm L}$,
\begin{eqnarray}
\label{BCST}
\frac{\widetilde{T}_c}{T_c} = 1 -\frac{2
b'_{T,k_0}\bar{\rho} (v-v_c^{\rm L})}{a_0 k_0 (1-{\chi}_T)}
\approx 1-\frac{v-v_c^{\rm L}}{v_{\rm F}}\,.
\end{eqnarray}
In the last equality we put $b'_{T,k_0}=b_0$. We also estimate the
maximal second critical velocity as $v_{c2}^{\rm max}\simeq
v_c^{\rm L} + v_{\rm F}$.

\begin{figure} %%%%% size was 3.8
\centerline{
\includegraphics[width=6cm]{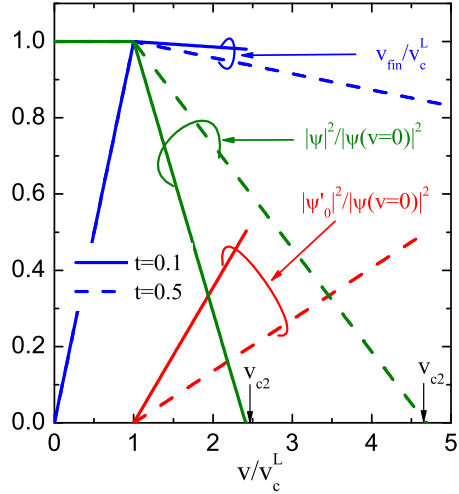}
}
\caption{Condensate amplitudes
$|\psi|^2$ and $|\psi'_0|^2$, Eq.~(\ref{orderp}), and the final
flow velocity $v_{\rm fin}$, Eq.~(\ref{vfin}), in superfluid
$^4$He plotted as functions of the flow velocity for various
temperatures. Vertical arrows indicate $v_{c2}$. Velocities are scaled by
the values of the Landau critical velocities $v_c^{\rm
L}(t=0.5)=59\,{\rm m/s}$ and $v_c^{\rm L}(t=0.1)=55\,{\rm m/s}$,
and the condensates are normalized to the condensate amplitude in
the superfluid at rest. } \label{fig:hecond}
\end{figure}

\subsection{bosonic superfluid on example of  He-II}

We turn now to the bosonic superfluid, He-II. In He-II there exists a branch of the phonon-roton excitations~\cite{LP1981,GS}.
The typical energy of the rotonic excitations $\Delta_{\rm r}=\epsilon (k_{\rm r})$ at the roton minimum  $k=k_{\rm r}$ depends on the pressure and temperature. According to~\cite{BrooksDonelly}, for the saturated vapor pressure $\Delta_{\rm r}= 8.71$\,K at $T= 0.1$\,K and $7.63$K  at $T=2.10$\,K, and $k_{\rm r} \simeq 1.9\cdot 10^{8}\hbar/{\rm cm}$ in the whole temperature interval.
%(We put the Boltzmann constant $k_{\rm B}=1$). Making use of the
%values of the parameters presented above and t
Other parameters of He-II at the saturated
vapor pressure are~\cite{GS}:
\begin{eqnarray}
&&T_c =2.17\,{\rm K},\,\,\,
a_0/ T_c^{4/3}=1.11\cdot 10^{-16}{\rm erg/K^{4/3}},\,\,\,
\nonumber\\
&&b_0/ T_c^{2/3} =3.54\cdot 10^{-39}{\rm erg\cdot cm^3/K^{2/3}}
\nonumber
\end{eqnarray}
and $c=c_0=1/m^{*}\simeq 1/m$,
with the helium atom mass  $m=6.6\cdot 10^{-24}$\,g. The
parametrization holds for $10^{-6}<t<0.1$, but for rough estimates
can be used up to $t=1$. For instance, using Eq.~(\ref{FGL}) we
evaluate the He-II mass-density as $m a_0/b_0\simeq 0.3\,{\rm
g/cm^{3}}$, which is of the order of the experimental value
$\rho_{\rm He}=0.15\,{\rm g/cm^{3}}$ at $P= 0$.

Taking into account
that we deal with the rotonic excitation, {\it i.e.}, $k_0\simeq
k_{\rm r}$  and $\epsilon(k_0)\simeq \Delta_{\rm r}$, we estimate,
$${k_0^2}/{(b_0\,\bar{\rho})}\simeq 47\,,\,\,
v_c^{\rm L}(T\to 0)\simeq 60\,{\rm {m}/{s}}\,, \,\,
 {a_0}/{k_0}\simeq 16{\rm {m}/{s}}.$$
Taking from~\cite{Pitaev84} that $b''_{T,k_0}\simeq 3.3
b_T$, and assuming $b'_{T,k_0}\sim b_T$ we again estimate
$0<{\chi}_T\ll 1$.
%%%%%%%%%%%%%%%%%%%%%%%%%%%%%%
Using the results of~\cite{BrooksDonelly}  $v_c^{\rm L}(T)$ dependence can be fitted with 99\% accuracy as
$$v_c^{\rm L}(T)/v_c^{\rm L}(0)\simeq
1-0.7e^{-2.14/\tilde{t}}+200\tilde{t} e^{-8/\tilde{t}}\,,$$ where
$\tilde{t}=T/T_c$.
%%%%%%%%%%%%%%%%%%%%%%%%%%%%%%
Using $\chi_T(\mbox{He-II})\sim \chi_T({\rm BCS})$, we evaluate condensate
amplitudes and the final flow velocity as functions of temperature
and depict them in Fig.~\ref{fig:hecond}. The CoE appears at
$v=v_c^{\rm L}$ in a second-order phase transition. For
$v>v_c^{\rm L}$ the amplitude of the condensate $|\psi'_0|^2$
($|\psi|^2$) increases (decreases) linearly with $v$. The closer
$T$ is to $T_c$, the steeper  the change of the condensate
amplitudes is. The final velocity of the flow, which sets in after
the appearance of the CoE, decreases with the increase of $v$.
With $\alpha =4/3$, $\beta =2/3$ the renormalized critical
temperature determined by Eq.~(\ref{tildet}) is
${\widetilde{T}_c}/{T_c}\approx 1 - 0.05\,(v/v_c^{\rm
L}(T_c)-1)^{3/2}$ for $v>v_c^{\rm L}$. The mother condensate
$|\psi|^2$ vanishes when $v$ reaches the value $v_{c2}$, which
depends on the temperature as $v_{c2}\approx v_c^{\rm L}(t) + (363
t^{2/3}-23.5t^{4/3}){\rm m/s}$. At $v=v_{c2}$ the superfluidity
disappears in a first-order phase transition. The corresponding
energy release can be estimated from~(\ref{jumpvc2}) as $ {\delta
F(v_{c2})}\approx \frac{47 a_0^2}{8b_0} t^{4/3}\simeq 5.9 \,
t^{4/3}{( T_c\Delta C_p)}$, where $\Delta C_p=0.76\cdot 10^7\,{\rm
erg/(cm^3 K)}$ is the specific heat jump at $T_c$~\cite{GS}.

\section{Rotating superfluids. Pulsars}\label{Pulsars}

The novel phase with the CoE may also exist
in rotating systems.  Here, excitations can be generated because
of the rotation. Now we should use the angular momentum conservation instead of the
momentum conservation. Also, the structure of the order parameter
is more complicated than the plane wave. For the cylindrical geometry a probing CoE function can be taken in the form~\cite{Voskresensky:1993uw}
\begin{eqnarray}
\psi'=\psi_0\exp\Big[ik_0 \tilde{r}\sin\Big(\phi-\alpha\frac{\om t}{k_0 \tilde{r}}\Big)-i\beta\om\,t\Big]\,,
\end{eqnarray}
where $\tilde{r}$ and $\phi$ are the polar coordinates and $\alpha$ and $\beta$ are variational parameters. The value of the critical angular velocity for the appearance of the first vortices, $\Omega_{c1} \sim v_{c1}/R$, proves to be very low for
systems of a large size $R$, e.g. like neutron stars. With these modifications, the
results, which we obtained above for the motion with the constant
$\vec{v}$, continue to hold.

In the inner crust and in a part of the core of a neutron star,
protons and neutrons are paired in the 1S$_0$ state owing to
attractive $pp$ and $nn$ interactions, cf. \cite{ST83}. In denser regions of the
star interior the 1S$_0$ pairing disappears but neutrons might be
paired in the 3P$_2$ state. The charged $pp$
superfluid component should co-rotate with the normal matter.
This, as we have mentioned,  is due to  the appearance of a tiny
magnetic field $\vec{h}=2m_p\vec{\Omega}/e_p$ (London effect) in
the whole volume of the superfluid, $m_p$ ($e_p$) is the proton
mass (charge)~\cite{Sauls89}. This tiny field, being $\lsim
10^{-2}$G for the most rapidly rotating pulsars, has no influence
on parameters of the star and can be neglected.

With the typical neutron star radius, $R\sim 10$\,km, and for
$\Delta\sim$MeV typical for the $1S_0$ $nn$ pairing, we estimate
$\Omega_{c1}\sim 10^{-14}$~Hz. For $\Omega\gg\Omega_{c1}$ the
neutron star contains arrays of neutron vortices with regions of
the superfluidity in between them, and the star rotates as a
rigid body. The vortices would completely overlap, only if
$\Omega$ reached unrealistically large value $\Omega_{c2}^{\rm
vort}\sim 10^{20}$~Hz.  The most rapidly rotating pulsar PSR
J1748-2446ad has the angular velocity
4500~Hz~\cite{Manchester:2004bp}. The value of the critical
angular velocity for the formation of the CoE in the neutron star
matter  is $\Omega_c\sim \Omega_c^{\rm L}\simeq \Delta/(p_{\rm
F}R)\sim 10^2$~Hz for the pairing gap $\Delta \sim$ MeV and
$p_{\rm F}\sim 300$\,MeV$/c$ at the nucleon density $n\sim n_0$,
where $n_0\simeq 0.17$fm$^{-3}$ is the density of the atomic nucleus, and $c$  is the
speed of light. The superfluidity will coexist with the
CoE and the array of vortices until the rotation frequency
$\Omega$ reaches the value $\Omega_{c2}>\Omega_c^{\rm L}$, at
which both the CoE and the superfluidity disappear completely.
From Eq.~(\ref{tildet}) with the BCS parameters we estimate
$\Omega_{c2}\sim v_{c2}/R\lsim 10^4$ Hz.

There are many other millisecond pulsars in low-mass X-ray binaries of a typical age $\gsim 10^8$\,yrs.
Thus, in the detected rapidly rotating pulsars the CoE might coexist with superfluidity,
that would also affect their hydrodynamical description~\cite{AndreevBashkin}.
A possible influence of the CoE on the window of the r-mode instability in the millisecond pulsars was recently studied by us in~\cite{KV2015}. Also a CoE may appear in the presence of a charged pion condensate with a finite momentum in massive neutron stars~\cite{Migdal:1990vm}, see a discussion of an additional slowing down of the pulsar which may  arise owing to the presence of the $\pi^+$ condensation  in~\cite{Voskresensky:1993uw}.  In massive neutron stars there may also exist  $K^-$ and/or $\bar{K}^0$ condensates
with a finite momentum, cf.~\cite{Kolomeitsev:1995xz,Kolomeitsev:2002pg}. A similar effect to that on a charged pion condensate may exist on $K^-$ and $\bar{K}^0$ condensates.
Another interesting issue is a possibility of the formation of CoEs in  color-superconducting regions of rotating hybrid stars. Various CoEs may arise there since pairing gaps between quarks of different colors and flavors may have essentially  different values, e.g. in 2SC, 2SC+X, color spin locking, and other possible phases, see in~\cite{BGV2001}.

\section{Conclusion}\label{Conclusion}

In this paper we studied a possibility of the condensation of excitations with $k\neq 0$, when a superfluid initially flows with respect to a wall with a velocity $v$ larger than the Landau critical velocity $v_c^{\rm L}$.
In difference with Refs.~\cite{Pitaev84,Ancilotto05,BP12}, which studied bosonic superfluid systems for $T=0$ at a fixed velocity $v$, we considered this phenomenon for bosonic and fermionic superfluid systems both for $T=0$ and $T\neq 0$ at the conserving momentum for a rectilinear
motion (at the conserving angular momentum for a rotation). In the presence of the CoE the final velocity of the superfluid   $v_{\rm fin}$ becomes less than $v$.
Also, compared to Refs.~\cite{Pitaev84,Voskresensky:1993uw,Ancilotto05} we incorporated  the interaction between the CoE and the ``mother'' condensate
of the superfluid. We studied the case of $T\ll T_c$, when the normal component can be neglected, and the case of higher $T$, when it serves as a reservoir of particles affecting the formation of the mother condensate and CoE. The latter case was not enlighten  yet in the literature.

At finite temperatures we first studied the systems where the superfluid and normal components  move with respect to each other with a relative velocity $\vec{w}$ (the average velocity of excitations with respect to the superfluid component), and then focused on the case of
$w=0$. Note that at finite $T$ the mother condensate may exist only for very low values of $\vec{w}$ (much less than the Landau critical velocity). In rotating superfluids vortices form a lattice and the system rotates as a rigid body. Also, charged subsystems are forced to move as a whole owing to a London force. These are conditions when indeed one can put $w=0$.

A back reaction of the CoE on the mother condensate
proves to be important both for $T=0$ and for $T\neq 0$. We found that the CoE
appears in a second-order phase transition at $v=v_c^{\rm L}$ and
the condensate amplitude grows linearly with the increasing
velocity. Simultaneously the mother condensate decreases and
vanishes at $v=v_{c2}$, then the superfluidity is destroyed in a
first-order phase transition with an energy release. For
$v_{c}^{\rm L}<v<v_{c2}$ the resulting flow velocity is $v_{\rm
fin}< v_c^{\rm L}$.

We found that
for the cold fermion systems with pairing the creation of the condensate of bosonic excitations with finite momentum, leading to a  reduction of the flow velocity up to the value of the Landau critical velocity $v_{c}^{\rm L}$, is energetically more profitable than the breaking of Cooper pairs appearing for $v>v_{c, {\rm f}}^{\rm L}$ ($v_{c}^{\rm L}>v_{c, {\rm f}}^{\rm L}$) and the decrease of the pairing gap (except the case when initial velocity $v$ is in a narrow vicinity of the critical point). To the best of our knowledge possibility of condensation of bosonic excitations with finite momentum in moving  fermionic systems with pairing was not yet considered in the literature.
  For fermion superfluids at $T\neq 0$   after the CoE is formed
the flow velocity becomes less than $v_{c,{\rm f}}^{\rm L}$ and  the Cooper pair
breaking does not occur, whereas the condensate of Bose
excitations is preserved. The CoE appears in the second-order phase transition. The mother condensate decreases and
vanishes at $v=v_{c2}(T)$, then the superfluidity is destroyed in a first-order phase transition with an energy release.

We discussed condensation of Bose excitations in rotating superfluids, such as pulsars and showed
that  in the existing most rapidly
rotating millisecond pulsars superfluidity might coexist with the CoE.\\[5mm]

We thank M.Yu.~Kagan for the detailed discussion of the results.
The work was supported  by the Ministry of
Education and Science of the Russian Federation (Basic part), by Slovak
Grant No. VEGA-1/0469/15, and by ``NewCompStar'', COST Action MP1304.

%========================================


\begin{thebibliography}{99}

\bibitem{Pitaev84}
L.P.~Pitaevskii,
%Layered structure of superfluid $^4$He with supercritical motion,
JETP Lett. {\bf 39}, 511 (1984).

\bibitem{Voskresensky:1993uw}
D.N.~Voskresensky,
%Condensate with a finite momentum in a moving medium,
JETP {\bf 77}, 917 (1993).

\bibitem{Vexp95}
D.N.~Voskresensky,
% ``Exponential growth and possible condensation of the particle-hole excitations in moving hot Fermi liquids'',
Phys. Lett. B {\bf 358}, 1 (1995).

%\cite{Kolomeitsev:2015pia}
\bibitem{Kolomeitsev:2015pia}
E.~E.~Kolomeitsev and D.~N.~Voskresensky,
%``Pomeranchuk instability and Bose condensation of scalar quanta in a Fermi liquid,''
arXiv:1505.03884 [nucl-th].

\bibitem{Ancilotto05}
F.~Ancilotto, F.~Dalfovo, L.P.~Pitaevskii, and F.~Toigo,
Phys. Rev. B {\bf 71}, 104530 (2005).

\bibitem{BP12}
G.~Baym and C.J.~Pethick,
%``Landau critical velocity in weakly interacting Bose gas'',
Phys. Rev. A {\bf 86}, 023602 (2012).

\bibitem{KV2015} E.E.~Kolomeitsev and D.N.~Voskresensky,
% Viscosity of neutron star matter and r-modes in rotating pulsars,
Phys. Rev. C {\bf 91}, 025805 (2015).

\bibitem{Melnikovsky}
L.A.~Melnikovsky,
%``Bose-Einstein condensation of rotons'',
Phys. Rev. B {\bf 84}, 024525 (2011).

\bibitem{LP1981}
E.M.~Lifshitz and L.P.~Pitaevskii
{\it Statistical Physics, Part 2}, Pergamon Press, Oxford, 1980.

\bibitem{GS}
V.L.~Ginsburg and A.A.~Sobyanin,
Sov. Phys. Usp. {\bf 19}, 773 (1977).

\bibitem{Khalatnikov}
I.M.~Khalatnikov,
{\it An Introduction to the Theory of Superfluidity}, Benjamin,  N.Y., 1965.

\bibitem{coldatmode}
I.~Shammass, S.~Rinott, A.~Berkovitz, R.~Schley, and J.~Steinhauer,
%``Phonon dispersion relation of an atomic Bose-Einstein condensate'',
Phys. Rev. Lett. {\bf 109}, 195301 (2012);
L.-Ch.~Ha, L.W.~Clark, C.V.~Parker, B.M.~Anderson, and Ch.~Chin,
%``Roton-maxon excitation spectrum of Bose condensates in a shaken optical lattice'' {\it arXiv:1407.7157}
Phys. Rev. Lett. {\bf 114}, 055301 (2015);
M.A.~Khamehchi, Y.~Zhang, C.~Hamner, Th.~Busch, and P.~Engels,
%``Long-range interactions and roton minimum softening in a spin-orbit coupled Bose-Einstein condensate'' {\it arXiv:1409.5387}
Phys. Rev. A {\bf 90}, 063624 (2014).

\bibitem{Kolomeitsev:1995xz}
E.~E.~Kolomeitsev, B.~K\"ampfer and D.~N.~Voskresensky,
%``Kaon polarization in nuclear matter,''
Nucl.\ Phys.\ A {\bf 588}, 889 (1995).

\bibitem{Kolomeitsev:2002pg}
E.~E.~Kolomeitsev and D.~N.~Voskresensky,
%``Negative kaons in dense baryonic matter,''
Phys.\ Rev.\ C {\bf 68}, 015803 (2003).

\bibitem{Migdal:1990vm}
A.B.~Migdal, E.E.~Saperstein, M.A.~Troitsky, and D.N.~Voskresensky,
%``Pion degrees of freedom in nuclear matter''
Phys. Rept.  {\bf 192}, 179 (1990).

\bibitem{ST83}
S.~Shapiro and S.A.~Teukolsky, {\it Black Holes, White Dwarfs and
Neutron Stars: The Physics of Compact Objects}, Wiley, N.Y., 1983.
\bibitem{Kagan}
R.~Combescot,  M.Yu.~Kagan, and S.~Stringari,
Phys. Rev. A {\bf 74}, 042717 (2006).

\bibitem{Urban}
N.~Martin and M.~Urban,
%``Collective Modes in a Superfluid Neutron Gas within the Quasiparticle Random-Phase Approximation,''
Phys. Rev. C {\bf 90}, 065805 (2014).
\bibitem{Bailin:1983bm}
D.~Bailin and A.~Love,
%``Superfluidity and Superconductivity in Relativistic Fermion Systems,''
Phys.\ Rept.\  {\bf 107}, 325 (1984).

\bibitem{Tilly-Tilly}
D.R.~Tilley and J.~Tilley, {\it Superfluidity and Superconductivity},
IoP Publishing, Bristol, 1990.

\bibitem{Abrikosov}
A.A.~Abrikosov, {\it Fundamentals of the Theory of Metals},
North Holland, Amsterdam, 1988.

\bibitem{GL}
V.L.~Ginzburg  and A.P.~Levanyuk, J. Phys. Chem. Solids {\bf 6}, 51 (1958);
A.P.~Levanyuk, Zh. Eksp. Teor. Fiz. {\bf 36}, 810 (1959)
[Sov.Phys. JETP {\bf 9}, 571 (1960)].

\bibitem{Amit68}
D.J.~Amit, Phys. Lett. A {\bf 26}, 466 (1968).

\bibitem{Kramer}
L.~Kramer, Phys. Rev. {\bf 179}, 149 (1969).

\bibitem{Andreev-Melnik04}
A.F.~Andreev and L.A.~Melnikovsky,
%``Two fluid model: new aspects'',
Physica C {\bf 404}, 34 (2004);
A.F.~Andreev and L.A.~Melnikovsky,
%``Thermodynamics of superfluidity'',
J. Low. Temp. Phys. {\bf 135}, 411 (2004).










\bibitem{Kulik43}
I.O.~Kulik, O.~Entin-Wohlman, and R.~Orbach
J. Low Temp. Phys. {\bf 43}, 591 (1981).

\bibitem{Castin14}
Y.~Castin, I.~Ferrier-Barbut, and C.~Salomon, Comptes Rendus--Phys. {\bf 16}, 241 (2015)
[arXiv:1408.1326].

\bibitem{Bardeen}
J.~Bardeen,
% Critical Fields and  Currents in Superconductors
Rev. Mod. Phys. {\bf 34}, 667 (1962).

\bibitem{Zagozkin}
A.M.~Zagoskin, {\it Quantum Theory of Many-body Systems},
Springer, N.Y., 1998, p.~182.

\bibitem{Lif81}
E.M.~Lifshiz and L.P.~Pitaevskii, {\it Physical Kinetics},
Pergamon Press, Oxford, 1981.


\bibitem{Sauls89}
J.~Sauls, {\it Timing Neutron Stars},
editors H.~\"Ogelman and E.P.J.~van den Heuvel,
Kluwer Academic Publishers, Dordrecht, 1989, pp.~457-490.


\bibitem{AndreevKagan84}
A.F.~Andreev and M.Yu.~Kagan,
%``Hydrodynamics of rotating superfluid liquid'',
Sov. Phys. JETP {\bf 59}, 318 (1984).

\bibitem{vortex-creation}
V.~Iordanskii, Sov. Phys. JETP {\bf 48}, 708 (1965);\\
J.S.~Langer and M.E.~Fisher, Phys. Rev. Lett. {\bf 19}, 560 (1967).

\bibitem{Tsuzuki}
T.~Tsuzuki, J. Low Temp. Phys. {\bf 4}, 441 (1971).

\bibitem{McClintock}
P.V.E.~McClintock and R.M. Bowley, {\it The Landau critical velocity}, Progress in Low Temp. Phys., Vol. XIV, Ed. by W.P. Halperin, Elsevier Sci. B.V., 1995.

\bibitem{BrooksDonelly}
J.S.~Brooks and J.~Donnelly,
J. Phys. Chem. {\bf 6}, 51 (1977).

\bibitem{Manchester:2004bp}
The ATNF Pulsar Catalogue, R.N.~Manchester, G.B.~Hobbs, A.~Teoh, and M.~Hobbs,
%``The Australia Telescope National Facility pulsar catalogue,''
Astronom. J. {\bf 129}, 1993 (2005);
Web address: \verb!http://www.atnf.csiro.au/research/pulsar/psrcat!

\bibitem{AndreevBashkin}
A.F.~Andreev and I.M.~Lifshitz,
%``Quantum Theory of Defects in Crystals,''
Sov. Phys. JETP {\bf 29}, 1107 (1969);
A.F.~Andreev and E.P.~Bashkin,
%``Three-velocity hydrodynamics of superfluid solutions'',
Sov. Phys. JETP {\bf 42}, 164 (1976);
E.P.~Bashkin and A.E.~Meyerovich,
Adv. Phys. {\bf 30}, 1 (1981).

\bibitem{BGV2001} D.~Blaschke, H.~Grigorian, and D.N.~Voskresensky,
%``Cooling of hybrid neutron stars and hypothetical selfbound objects with superconducting %quark cores'',
Astron. Astrophys. {\bf 368}, 561 (2001).










\end{thebibliography}
\end{document}